# A spin Esaki diode

Makoto Kohda, Yuzo Ohno, Koji Takamura, Fumihiro Matsukura, and Hideo Ohno

Laboratory for Electronic Intelligent Systems, Research Institute of Electrical Communication, Tohoku University, 2-1-1 Katahira, Aoba-ku, Sendai 980-8577 Japan

Abstract

We demonstrate electrical electron spin injection via interband tunneling in ferromagnetic/nonmagnetic semiconductor Esaki diodes. An interband tunnel junction between ferromagnetic $p^+$-(Ga,Mn)As and nonmagnetic $n^+$-GaAs under reverse-bias allows spin-polarized tunneling of electrons from the valence band of (Ga,Mn)As to the conduction band of $n^+$-GaAs. The spin polarization of tunneled electrons is probed by circular polarization of electroluminescence (EL) from an n-GaAs/InGaAs/p-GaAs light emitting structure integrated with the diode. Clear hysteresis loop with ± 6.5% remanence is observed in the magnetic-field dependence of the EL polarization at 6 K, below the Curie temperature of (Ga,Mn)As.

Semiconductor spin-electronics (spintronics), where both charge and spin degrees of freedom are utilized, is currently of great interest. This is because it is expected to enable integration of magnetic functionalities and semiconductor circuitry as well as quantum information technology based on coherent nature of spins in semiconductors[1]. To put these devices into practice, generation and injection of spin polarized carriers into a non-magnetic semiconductor are indispensable. Recently electrical spin injection has been demonstrated with ferromagnetic semiconductor (Ga,Mn)As[2], paramagnetic semiconductor (Be)ZnMnSe[3,4] and ferromagnetic metal Fe[5].

For a number of spin-based device schemes, electrical spin injection through high quality semiconductor heterojunctions in the absence of magnetic field is preferable. At the same time, electron spin injection is desired as electrons provide two-level system and smaller spin-orbit coupling compared to holes. The known ferromagnetic semiconductors compatible with high quality heterostructures are, however, all p-type[6,7]. This is believed to be due to the small exchange interaction between magnetic spins and conduction band electron spins[6,7]. Here we demonstrate electrical *electron* spin injection into non-magnetic semiconductors from a *p-type* ferromagnetic semiconductor using interband tunneling (a spin Esaki diode[8]).

Our scheme is schematically shown in Fig. 1 (a). An interband tunnel junction with ferromagnetic $p^+$-(Ga,Mn)As[9,10] and nonmagnetic $n^+$-GaAs is epitaxially grown on top of a non-magnetic n-GaAs/(In,Ga)As/p-GaAs quantum well light emitting diode (LED) structure. The LED is used to probe the electrical electron spin injection by analyzing the relative circular polarization of electroluminescence ($\Delta P_{EL}$). When a bias is applied to the whole structure in such a way that the nonmagnetic pn junction is forward biased, a small reverse bias appears at the $p^+$-(Ga,Mn)As/$n^+$-GaAs tunnel junction and spin polarized electrons are injected into the $n^+$-GaAs/n-GaAs part of the structure. These spin polarized electrons reach the (In,Ga)As quantum well (QW) and recombine with spin unpolarized holes supplied from the bottom p-GaAs. The electron spin polarization results in a circularly-polarized light emission, which is used to detect the polarization of the injected electrons.

We first examined the current-voltage (*I-V*) characteristics of an $p^+$-(Ga,Mn)As/$n^+$-GaAs single pn junction. Figure 1 (b) shows the *I-V* characteristics of a diode having 30 μm-square mesa of 200 nm $Ga_{0.963}Mn_{0.037}As/n^+$-GaAs ($N_D = 1 \times 10^{19}$

cm$^{-3}$) grown on an n-GaAs substrate measured at 77 K. When forward bias is applied (positive on (Ga,Mn)As), the conduction is suppressed and a negative differential resistance is observed in the range $V$ = 260 - 370 mV, a clear evidence of an Esaki diode operation. When a reverse bias is applied, high conductivity was achieved at small voltages, indicating tunneling of the valence electrons in p$^+$-(Ga,Mn)As to the conduction band of n$^+$-GaAs.

We then added an LED structure below the Esaki diode, in order to probe electrical electron spin injection through $\Delta P_{EL}$. The nonmagnetic LED structure was first grown in a molecular beam epitaxy (MBE) system dedicated for non-magnetic materials (no magnetic source is installed), followed by an overgrowth of ferromagnetic (Ga,Mn)As in a separate MBE system equipped with an Mn source[11]. The LED structure consists of, from the top, 10 nm n$^+$-GaAs (doping density $N_D \sim 2\times10^{19}$ cm$^{-3}$), 200 nm n-GaAs ($N_D = 10^{17}$ cm$^{-3}$), an undoped active layer (10 nm In$_y$Ga$_{1-y}$As with $y$ = 0.08 - 0.13 sandwiched by 10 nm GaAs), 100 nm p-GaAs (acceptor density $N_A = 10^{17}$ cm$^{-3}$), and a p$^+$-GaAs ($N_A = 2\times10^{18}$ cm$^{-3}$) buffer layer. To avoid undesired spin scattering, we reduced $N_D$ in the spin-transport layer (200 nm n-GaAs)[12]. The thickness of the thin n$^+$-GaAs layer was chosen to introduce donors necessary to prevent depletion of n-GaAs. After the growth of the nonmagnetic part, the wafer was capped with arsenic for surface protection and then transferred to the other MBE chamber where a 300 nm (Ga$_{1-x}$,Mn$_x$)As with $x$ = 0.035 ± 0.003 was overgrown at 250 °C. The grown structure was processed into 200 μm-wide mesa stripes with a top metal electrode (Au/Cr). For edge emission, the wafer was cleaved into 1 mm-long pieces, soldered on a circuit board, and then wired. Figure 1 (c) shows the $I$-$V$ characteristics of thus prepared ferromagnetic tunnel junction with an LED. Here the positive bias is defined with respect to a non-magnetic pn junction. Figure 2 shows the EL spectra of the Esaki diode/LED structure measured at 4.5 K (edge emission). We observed an intense peak at $E$ = 1.39 eV from the (In,Ga)As quantum well with the full-width at half-maximum = 7 meV. No emission peak from the bulk region of GaAs was observed.

To examine the electron spin injection from the ferromagnetic p$^+$-(Ga,Mn)As layer into n-GaAs via interband tunneling, we measured circular polarization of electroluminescence in an applied magnetic field. The Esaki diode/LED structure was placed in a magneto-optical cryostat in a Faraday configuration. The LED was driven by pulsed current at 671 Hz. The emitted light was analyzed by a Babinet-Soleil phase compensator (used as a variable $\lambda/4$ waveplate) and a linear-polarizing beam splitter, which separate right-($\sigma^+$) and left-($\sigma^-$)-circularly polarized components $I^{\sigma\pm}$. Each signal was detected by a silicon pin photo-diode. The degree of polarization was defined by $P_{EL} = (I^{\sigma+}-I^{\sigma-})/(I^{\sigma+}+I^{\sigma-})$.

In Fig. 3 (a), traces of relative polarization of EL $\Delta P_{EL}$ are displayed as a function of in-plane magnetic field for temperatures between 6 and 50 K (edge emission). The magnetic field was swept from 1 T to –1 T. $\Delta P_{EL}$ shows hysteresis with loops and the coercive field decreases from 5 to 0 mT as the temperature increases from 6 K to near $T_C$ (= 60 K). The maximum $\Delta P_{EL}$ is ±6.5 % at 6 K. We observed almost no dependence of the polarization caused on the injection current level from 1 to 30 mA.

We also investigated the polarization of light emitted from the QW plane. In this configuration, we can directly relate $\Delta P_{EL}$ to the spin polarization of electrons by taking into account the selection rule for optical transition between quantized states in the quantum well. To allow measurements of surface emitted light, we polished the backside of the substrate, mounted on a fused silica glass plate, and fixed it in a magneto-optical cryostat so that a magnetic field can be applied normal to the QW plane. Figure 3 (b) displays $\Delta P_{EL}$ at 8 K. Since the magnetic field direction is now normal to the magnetic easy axis of (Ga,Mn)As (hard axis), no hysteresis is observed in $\Delta P_{EL}$. $\Delta P_{EL}$ at 1 T reaches 8.5 %, almost the same value as the $\Delta P_{EL}$ taken in the edge-emission configuration.

We have also fabricated a hole spin injection structure (a pn junction with an LED, see ref. 2) and compared the results with that of the present electron spin injection structure. The device structure for hole spin injection was prepared on a n-GaAs substrate. An active region consists of a 10-nm-thick $In_{0.15}Ga_{0.85}As$ QW sandwiched between 20-nm-thick undoped GaAs barrier layers. On top of this active structure, a 210-nm-thick i-GaAs spacer and a 1 μm-thick (Ga,Mn)As layer were grown. The device structure for the electron spin injection is the same as the device shown in Fig. 1 (c), except the (Ga,Mn)As thickness and indium concentration of QW; (Ga,Mn)As is 1 μm and the indium concentration of InGaAs QW is 15%, to match those used in the hole injection device. To minimize the run to run variation, the (Ga,Mn)As layer of the two devices are grown simultaneously in one run. The spacing between the (Ga,Mn)As layer and the QW are made virtually identical; 230 nm for the hole injection device and 220 nm for the electron injection device. After processed into 1-mm devices, the samples were placed in a magneto-optical cryostat and a magnetic field was applied parallel to the plane of the samples. $\Delta P_{EL}$ is shown in Fig. 4 as a function of in-plane magnetic field at 8K. $\Delta P_{EL}$ of the hole spin injection device is 0.8%, in accordance with the previous study[2], whereas $\Delta P_{EL}$ of the electron spin injection device is 3.8%; about 5 times larger than hole spin injection. This large difference excludes the magneto-optical effect of (Ga,Mn)As as the origin of the polarization, because the QW to (Ga,Mn)As separation is virtually the same; we expect that the polarization of the two samples is expected to be the same if only the magneto-optical effect from (Ga,Mn)As[13] were the origin. This observation indicates the origin of $\Delta P_{EL}$ being electrical spin injection.

In conclusion, we have demonstrated electrical electron spin injection from p-type ferromagnetic semiconductor into non-magnetic semiconductor using an Esaki diode structure. The electrical electron spin injection is demonstrated by circular polarization of light emitted from an integrated LED, which indicates that the carriers undergo interband tunneling with their spin polarization preserved. Further optimization of the device structure may enhance the spin injection efficiency further.

This work was partly supported by the "Research for the Future" program (No. JSPS-RFTF97P00202) from JSPS and by a Grant-in-Aids from the Ministry of Education, Culture, Sports, Science, and Technology, Japan (No. 12305001).

Figure captions

Fig. 1. (a) Schematic band structure of an Esaki diode/LED electrical electron spin injection device. An interband tunnel junction (Esaki diode) with ferromagnetic $p^+$-(Ga,Mn)As and nonmagnetic $n^+$-GaAs is epitaxially grown on top of a non-magnetic n-GaAs/(In,Ga)As/p-GaAs quantum well (QW) LED structure. Spin polarized electrons injected from $p^+$-(Ga,Mn)As into n-GaAs spacer reach the (In,Ga)As QW and recombine with spin unpolarized holes supplied from the bottom p-GaAs. (b) Current-voltage (*I-V*) characteristics of an $p^+$-(Ga,Mn)As/$n^+$-GaAs single pn junction with 200-nm (Ga,Mn)As and $n^+$-GaAs. (c) Current-voltage (*I-V*) characteristics of an Esaki diode/LED electrical electron spin injection device at 77 K.

Fig. 2. Electroluminescence (EL) spectrum of an Esaki diode/LED structure measured at 4.5 K. EL peak at $E = 1.39$ eV is from the (In,Ga)As quantum well. No emission peak from the bulk region of GaAs is observed.

Fig. 3. (a) Relative polarization of EL $\Delta P_{EL}$ as a function of in-plane magnetic field *B* for temperatures between 6 and 50K (edge emission). $V = 1.7$ V and $I = 20$ mA. (b) $\Delta P_{EL}$ as a function of magnetic field *B* perpendicular to the sample plane at 8 K (surface emission from the backside of the wafer). $V = 1.6$ V and $I = 9$ mA.

Fig. 4. The EL polarization of electron (solid square) and hole (open square) spin injection devices as a function of in-plane magnetic field at 8K(edge emission). $V = 1.9$ V and $I = 22$ mA for electron spin injection. $V = 3.4$ V, $I = 20$ mA for hole spin injection. $\Delta P_{EL}$ at $B = 0$ for electrons is 3.8%, while $\Delta P_{EL}$ for holes is 0.8%.

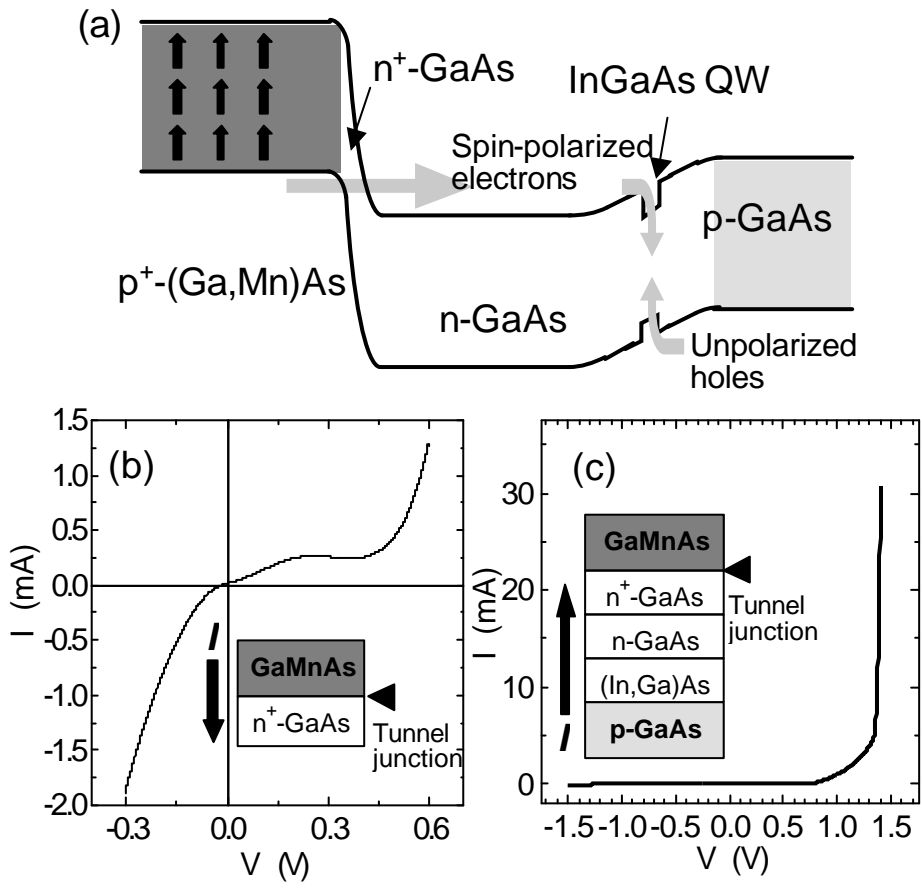

Fig. 1 Kohda et al.

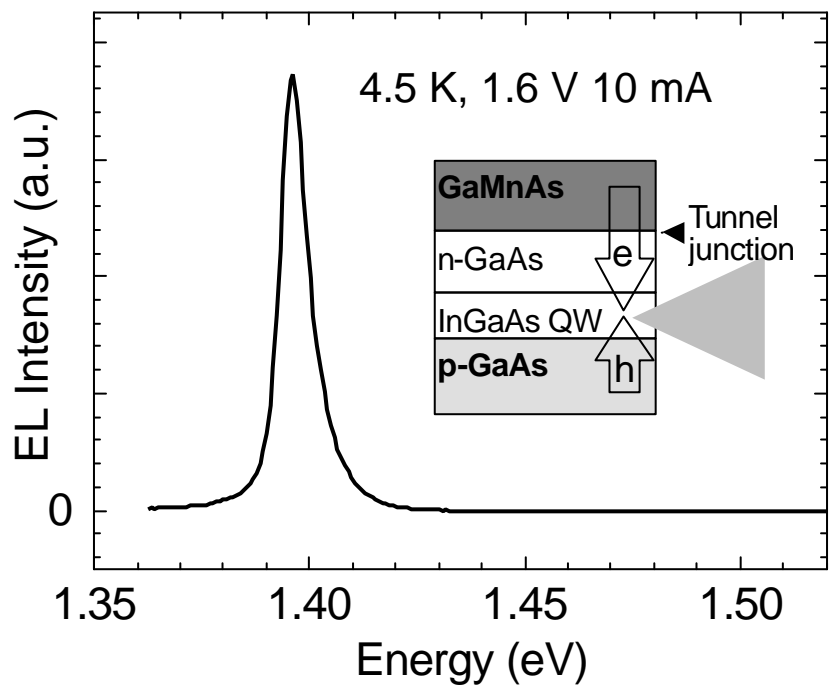

Fig. 2 Kohda et al.

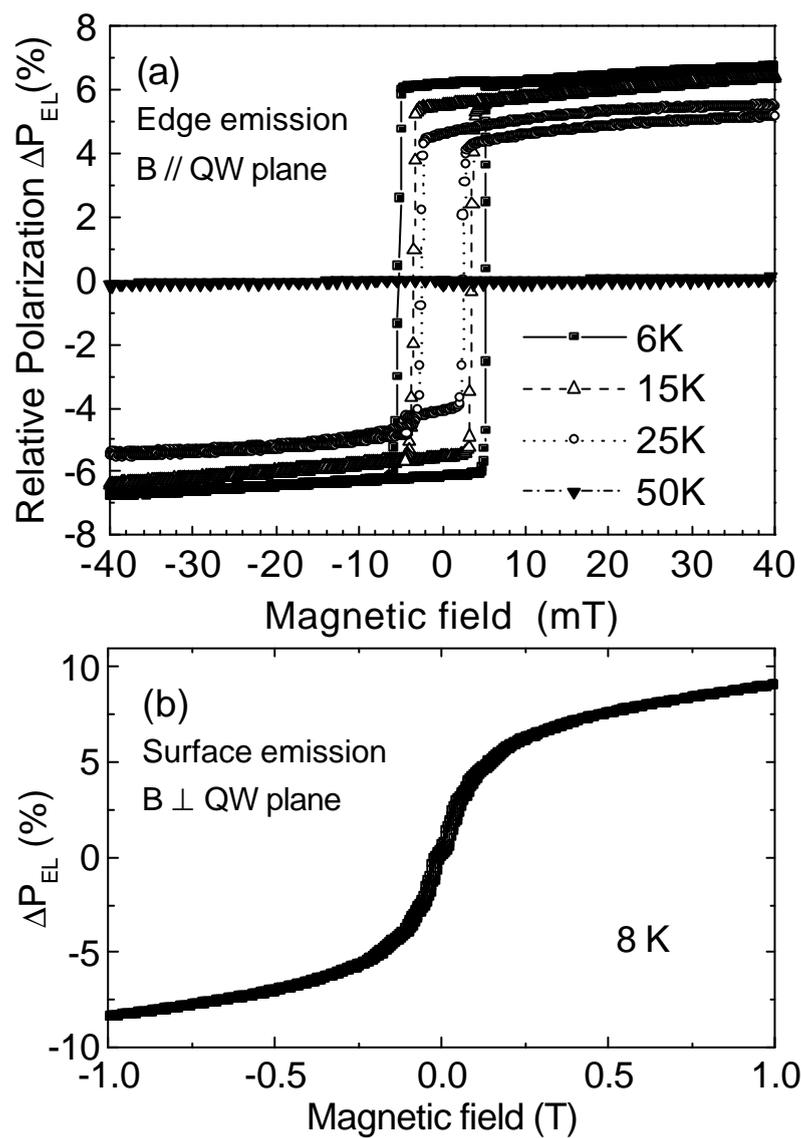

Fig. 3 Kohda et al.

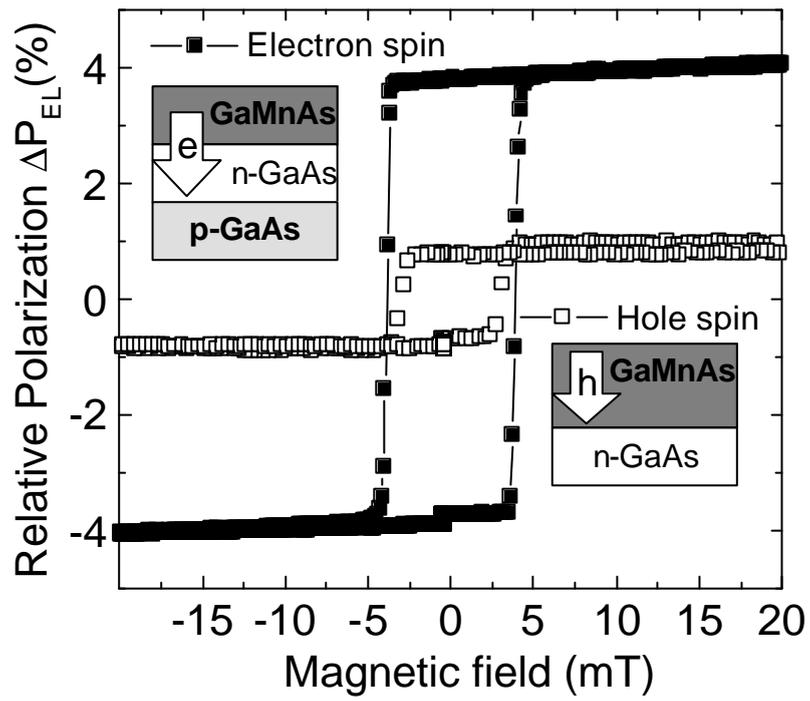

Fig. 4 Kohda et al.